\documentclass[12pt]{article}
\UseRawInputEncoding
\usepackage[hmargin=2.5cm,vmargin=3cm]{geometry}
 \usepackage{graphicx}
 \usepackage{amsmath}
 \usepackage{amssymb}
  \usepackage{enumerate}
\usepackage{cite}
\usepackage{mathabx}
\newcommand{\bey}{\begin{eqnarray}}
\newcommand{\eey}{\end{eqnarray}}

\begin{document}

\title{Gravity, Quantum Fields and Quantum Information:\\ Problems with classical channel and stochastic theories}
\author{C. Anastopoulos and B. L. Hu}

\date{{\small (Feb 4, 2022)}}
\maketitle

\begin{abstract}
In recent years an increasing number of papers  attempt to mimic or supplant  quantum field theory in discussions of issues related to gravity  by the tools and through the perspective of quantum information theory, often in the context of alternative quantum theories.  In this essay, we point out three common mistakes or inconsistencies in such treatments. First, we show that the notion of  interactions mediated by an information channel is not, in general, equivalent to the treatment of interactions by quantum field theory. When used to describe gravity, this notion may lead to inconsistencies with general relativity. Second, we point out that in general one cannot replace  a quantum field by that of classical stochastic sources,  or mock up the effects of quantum fluctuations by classical noises, because in so doing important quantum features such as coherence and entanglement will be left out.  Third, we explain how under specific conditions semi-classical and stochastic theories indeed can be formulated from their quantum origins and play a role at certain regimes of interest. \\


-- Prepared for Entropy: Special Issue on Quantum Information in Quantum Gravity.

\end{abstract}

\newpage
The rapid growth of quantum information science has led to the application of Quantum Information Theory (QIT) concepts  to different branches of physics, from condensed matter to cosmology and quantum gravity. In particular, QIT concepts are used to formulate models for  new physics, such as  alternative quantum theories (AQT) or  gravitational interactions. The key point we make in this paper is that these models must be structurally compared with our current best theories for the phenomena they purport to describe, namely, Quantum Field Theory (QFT) and General Relativity (GR). Even if these models have good motivations and their own logic, they must past the scrutiny of these two well-tested theories or be embedded in a framework that relies on these proven theories. Failure to do so renders  such models physically inconsistent and, hence, {\em a priori} implausible.

In this paper, we focus on three particular problems that commonly appear in such models. First, we show that the QIT notion of an information channel does not, in general,  capture the idea of mediating fields in QFT. It is therefore misleading to use it as a basis for describing, for example, the gravitational interaction. Second, we argue that we cannot in general substitute quantum fluctuations in an unknown theory with classical stochastic noises. Third, we explain the proper way to derive the stochastic and semi-classical theories from their quantum origins.

\section{Problem 1:  Quantum fields misconstrued as information channels}

\subsection{Quantum channels}

 A quantum channel is a map $C$ that takes  quantum states $\hat{\rho}_e$ on a Hilbert space ${\cal H}_A$ (emitter) to quantum states $C[\hat{\rho}_e]$ on a Hilbert space ${\cal H}_B$ (receiver). The map is completely positive and trace preserving.

 Two types of quantum transmission are directly relevant: Direct interaction channels and channels with a mediator.

\smallskip

\noindent $\bullet$ In a {\em direct interaction} channel, the emitter and receiver interact through a unitary map $\hat{U}$ on ${\cal H}_A\otimes {\cal H}_B$. Then
\bey
C[\hat{\rho}_e] = Tr_{{\cal H}_A}\left(\hat{U}(\hat{\rho}_e \otimes \hat{\rho}_0)\hat{U}^{\dagger}\right),
\eey
where $\hat{\rho}_0$ is the initial state of the receiver.  Usually $\hat{U}$ is generated by a Hamiltonian $\hat{H}_A  + \hat{H}_B + \hat{H}_I$, where the interaction term $\hat{H}_I$ is switched on for a finite time interval.

\smallskip

\noindent $\bullet$ In a {\em mediated interaction} channel, there exists a mediating system described by the Hilbert space ${\cal H}_M$. Both emitter and receiver interact through the mediator, but not directly with each other. Then, the Hamiltonian is $\hat{H} = \hat{H}_A + \hat{H}_B + \hat{H}_M + \hat{H}_{AM} + \hat{H}_{BM}$, where $\hat{H}_{AM}$ describes the interaction of the emitter with the mediator and $\hat{H}_{BM}$ describes the interaction of the receiver with the mediator. These interactions are usually viewed as permanent, i.e., that they cannot be switched on or off.

If $\hat{U}$ is the unitary evolution generated by this Hamiltonian, then
\bey
C[\hat{\rho}_e] = Tr_{{\cal H}_A\otimes {\cal H}_M}\left(\hat{U}(\hat{\rho}_e \otimes \hat{\rho}_0 \otimes \hat{\rho}_M)\hat{U}^{\dagger}\right),
\eey
where now $\hat{\rho}_M$ is the initial state of the mediator, typically taken as the ground state of the mediator's self-Hamiltonian $\hat{H}_M$.

\medskip

Emitter and receiver are often spatially localized and separated by a finite distance $L$, in which case the issue of signal propagation arises. Direct interactions lead to faster than light-signal, while it is expected that mediated transmission through local interactions respects causality.

\subsection{Problems with treating QFT as quantum channel}

Quantum communication protocols with photons often employ the electromagnetic (EM) field as a transmission channel. For example, telecommunications from radios to mobile phones treat the classical  EM  field as a channel for the transmission of information.   In these cases the classical EM field configurations are identified with specific states of the quantum EM field.   Therefore it is tempting to view quantum fields that mediate an interaction as defining informational channels.   However, the  language of channels does not capture the full physics of interacting quantum field theories (QFT). The reasons are the following.

\medskip

\noindent $\bullet$ A consistent theory of interactions exists only if receiver and emitter are also treated by QFT. But then, the Hamiltonian of the system cannot be defined on a tensor product Hilbert space ${\cal H}_A \otimes {\cal H}_B \otimes {\cal H}_{med}$, as in a mediated interaction channel. This statement follows from classic theorems by Haag \cite{Haag} and Hall and Wightman \cite{HaWi}. This means that
{\it the degrees of freedom of the field are intertwined with those of the emitter and the receiver,} in a way that it is impossible to disentangle. The root of this problem is phenomena like vacuum polarization -- it is impossible to describe the field vacuum as a factorized state, or even an entangled state in a factorized single state.

This implies that me must  use dressed states for the emitter and the receiver, i.e., the mediator degrees of freedom affect the available states of receiver and emitter. The sharp separation is possible only if we work to lowest order in perturbation theory, thus avoiding all Feynman diagrams with loops that mix the degrees of freedom of the emitter/receiver with those of the mediating field. The treatment of the field as an information channel does not make sense beyond this regime.

\medskip

\noindent $\bullet$ The notion of a quantum channel originates from Quantum Information Theory  (QIT), which has mainly been  developed in the context of non-relativistic quantum mechanics, a small corner of full QFT.  {\it Current QIT is problematic when basic relativistic principles – both special and general -- such as causality and covariance, need be accounted for}. A relativistic QIT that  expresses all informational notions in terms of quantum fields is currently missing, largely because of difficulties in formulating a comprehensive QFT theory of measurement \cite{AnSav22}.

 Here, we point out one particular problem, namely, the fact that the  notion of a localized quantum system in QFT leads to conflicts with causality. This is important  for the consistent definition of localized  receivers and emitters in QFT.   This result is evidenced by a number of theorems, for example, by Malament  \cite{Malament},  Schlieder \cite{Schlieder} and Hegerfeld \cite{Heg1}. These works show that the natural definitions of localized observables  together with Poincar\'e covariance and energy positivity  conflict   the requirement of relativistic causality.

The most well known set-up where localization appears to contradict causality is {\it Fermi's two-atom problem}. In the first analysis of a quantum field as a mediated interaction channel, Fermi analysed the interaction of two remote atoms through the quantum EM field  \cite{Fermi}. He  assumed that at time $t = 0$,  atom A (emitter) is in an excited state and atom B (receiver) in the ground state. He asked when B will notice a signal from A and leave the ground state. In accordance with Einstein locality, he found that this happens only at time greater than $r$. It took about thirty years for Shirokov to point out that Fermi's result is an artifact of an approximation \cite{Shiro}.
Several studies followed with conclusions depending on the approximations used. Eventually, Hegerfeldt showed that non-causality is generic \cite{Heg,Heg2}, as it depends solely on  energy positivity and on the treatment of atoms as localized in disjoint spatial regions.

\medskip

\noindent $\bullet$ {\it The naive idea of a field as an object that mediates interaction is insufficient} to describe the actual theories of mediating interactions, namely, gauge field theories. The reason is that it does not take into account the presence of constraints, which are a consequence of the gauge symmetry. The same issue appears in the treatment of gravity.

In classical field theory, a constraint means that some degrees of freedom that appear in the Lagrangian formulation are either slaved to other degrees of freedom, or pure gauge in the sense that they do not affect the properties of the system. In other words, they are not true degrees of freedom that propagate through the equations of motion, and they do not represent physical observables.

When quantizing a constrained field theory, probabilities must be expressed solely in terms of true degrees of freedom. All major approaches to quantization end up with a Hilbert space ${\cal H}_{phys}$, where true degrees of freedom live. The state of the field is given by vectors / density matrices on ${\cal H}_{phys}$.

For example, in the Hamiltonian treatment of the electromagnetic field, the EM potential can be split in four components: the scalar part $A_0$, the longitudinal component $A_i^L$ and the transverse component $A_i^T$. The conjugate of $A_0$ vanishes, while the conjugate $E_i^L$ of $A_i^L$ is slaved to the matter charge density $\rho$  through the Gauss constraint ${\bf \nabla} \cdot {\bf E} = \rho$. Then, $A_0$ and $A_i^L$ are pure gauge variables, and the true degrees of freedom consist of $A_i^T$ and their conjugate momenta $E_i^T$. The Hamiltonian for charged matter interacting with the EM field is
\bey
H = H_{mat} + H_{EM} + \frac{1}{2} \int d^3 x d^3 x' \frac{\rho({\bf x}) \rho({\bf x}')}{|{\bf x} - {\bf x}'|} +     \int d^3x {\bf j}\cdot {\bf A}^T , \label{htot}
\eey
where $H_{mat}$ is the self-Hamiltonian for matter, $H_{EM} $ the self-Hamiltonian of the EM field, and ${\bf j}$ is the current  density.
The last term, the Coulomb interaction, follows from the Gauss law constraint. It is non-local, and it depends solely on the matter degrees of freedom through the charge density $\rho$.

When quantizing all variables in Eq. (\ref{htot}) are promoted to operators. In a perturbative treatment of the EM interaction, we employ the  Hilbert space ${\cal H}_{mat}\otimes {\cal H}_{EM}$, where ${\cal H}_{mat}$ describes  the matter degrees of freedom and ${\cal H}_{EM}$ describes the field degrees of freedom. On this Hilbert space a Hamiltonian that  includes all terms in Eq.(\ref{htot}) but the last is in principle well defined. The last term describes the interaction of photons with matter, and it can be implemented perturbatively.  It is important to emphasize that the Coulomb term is defined as an operator of ${\cal H}_{mat}$, since it depends only on the charge density operator $\hat{\rho}({\bf x})$. The latter is well defined modulo proper regularization.

Hence, when splitting the matter  Hilbert space  ${\cal H}_{mat}$ as ${\cal H}_A \otimes {\cal H}_B$ in terms of the emitter $A$ and the receiver $B$,
we obtain
\bey
\hat{H} = \hat{H}_A + \hat{H}_B + \hat{H}_{EM} + \int d^3x \hat{\bf j}_A\cdot \hat{\bf A}^T+\int d^3x \hat{\bf j}_B \cdot \hat{\bf A}^T  + \int d^3 x d^3 x' \frac{\hat{\rho}_B({\bf x}) \hat{\rho}_A({\bf x}')}{|{\bf x} - {\bf x}'|} .
\eey
This means that the Hamiltonian contains an interaction of the two subsystems by the two terms: the  Coulomb term that causes direct non-local interaction between the two subsystems, and the interaction that is mediated by photons. It may be tempting to identify the former with a {\em direct channel} and the latter with a mediated channel, except for the fact that the former term cannot be switched off. The Coulomb interaction between the subsystems is always present; its expectation is non-zero even in the absence of photons. The Coulomb term affects the preparation of the system, and it may not allow for the preparation of an initial state that is uncorrelated. The presence of the Coulomb term is a consequence of the gauge symmetry of EM: it leads to constraints that define {\em instantaneous} laws in contrast to the causal dynamical laws of time evolution of the true degrees of freedom of the system.

\medskip

For the reasons above, we believe that the treatment of relativistic interactions in the language of information channels is justified only as  an approximation for specific tasks. Otherwise, it should be taken as a simile with a restricted domain of validity, and certainly not as an expression of fundamental physics.

\subsection{Implications for gravity}
The analysis for the EM field given previously can also be carried out for gravity in the weak coupling limit, where we treat the gravitational field as small perturbations around some background geometry, the simplest being  the Minkowski spacetime. The Hamiltonian in the Arnowitt-Deser-Misner gauge is  a sum of three terms  \cite{ADM, ALS}
\bey
H_{ADM} = H_{matt} +    H_{gsi}+ H_{GW} , \label{hadm}
\eey
where $H_{GW}$  is the Hamiltonian for gravitational waves, including a term for the interaction of gravitational waves with matter, and $H_{gsi}$ is a non-local interaction term ("gsi" stands for gravitational self-interaction). In the non-relativistic regime
\bey
H_{gsi} = -\frac{1}{2} \int d^3 x d^3 x' \frac{\mu({\bf x}) \mu({\bf x}')}{|{\bf x} - {\bf x}'|}
\eey
where $\mu({\bf x})$ is the mass density. The self-interaction term $H_{gsi}$ originates from the Poisson  equation for the gravitational potential
\bey
\nabla^2 \phi = - 4 \pi\mu, \label{poisson}
\eey
which is obtained as a {\em constraint} in the Hamiltonian analysis of the weak-field, non-relativistic limit of GR. The gravitational potential $\phi$ is a specific component of the three-metric tensor. Eq. (\ref{poisson}) implies that $\phi$ is not an independent variable in the Hamiltonian theory, rather it is slaved to the mass distribution.

The usual quantization procedure for weak gravity is to quantize the mass degrees of freedom separately from the gravitational degrees of freedom (gravitons), and to introduce the coupling through perturbation theory. The resulting theory should be finite at tree level, and work as an effective QFT for graviton physics. This procedure presupposes that quantization commutes with the weak-gravity limit. This is a plausible but  highly-non trivial assumption of perturbative quantum gravity.  If the two operations do not commute, the gravitational part of (\ref{hadm}) should not be treated as     an effective QFT with a well defined graviton vacuum.

The key point here is that the nature of the term $H_{gsi}$ is not affected by quantization hypotheses about the gravitational field, because it is a functional solely of the matter degrees of freedom. If matter is quantized, then the matter Hamiltonian    must include $\hat{H}_{gsi}$, because the latter is expressed in terms the mass density for matter. Then, Eq. (\ref{poisson}) holds at the level of operators,
\bey
\nabla^2 \hat{\phi} = - 4 \pi \hat{\mu}, \label{poisson}
\eey
 where $\hat{\phi}({\bf x})$ is an auxilliary operator constructed from the quantum mass distribution $\hat{\mu}({\bf x})$.
Obviously, if we purport to view the Newtonian  interaction between two remote masses given by $H_{gsi}$ as a channel, this channel will be direct.

Many authors have tried to view gravity as a fundamentally classical channel. This is only possible if Eq. (\ref{poisson}) fails to hold for a quantum mass distribution $\hat{\mu}({\bf x})$. Then, $\phi$ is a classical variable that is only partially correlated with the mass density. For example, in the alternative quantum theories based on the Newton-Schr\"odinger equation, the Poisson equation takes the form $\nabla^2 \phi = - 4 \pi\langle \hat \mu\rangle$, where the expectation value is taken with respect to the quantum state. Then, the interaction Hamiltonian of the system is
\bey
H_{gsi} = -  \int d^3 x d^3 x' \frac{\mu({\bf x}) \langle\hat{\mu}({\bf x}')\rangle}{|{\bf x} - {\bf x}'|},
\eey
i.e., it depends on the quantum state. This leads to a non-linear equation for a wave function, similar to Hartree's equation governing a large number of particles, which is often invoked in atomic and nuclear physics.  (Strictly speaking, the Hartree wavefunction  is  not a quantum state of the $N$-particle system, but a variational parameter in the mean-field approximation---see, for example, Ref. \cite{Schwabl}.)
The pitfall is to assert that this evolution equation  holds even for single or a few individual particles, as many proponents of the Newton-Schr\"odinger equation  do \cite{AHNSE,Adler}.

In this approach, Newtonian gravity is not a direct  interaction channel. The potential is essentially a stochastic variable correlated with the mass density $\mu_1$ of one subsystem;  this potential generates  a classical stochastic force for the other subsystem. This idea is made explicit by the Kafri-Taylor-Milburn (KTM) model \cite{KTM} and its significant upgrade by Diosi and Tilloy (DT) \cite{DioTi}, where gravity is modeled by a classical noisy channel. Effectively, this model amounts to the continuous measurement of the mass density $\mu_1$ of one subsystem\footnote{The original KTM model involves a continuous measurement of particle positions, and mainly focuses on deviations from an equilibrium configuration. The TD model considers the measurement of the mass density, and as such it provides a broader generalization of KTM's main concept. However, they are distinct models, as the restriction of the TD model to the regime where the KTM model applies gives different results \cite{GCB21}. }, and a classical stochastic measurement record $J_1$ that acts as a classical control force on the other mass. Again, $\phi$ in Eq. (\ref{poisson}) is treated as a stochastic classical  variable that  is correlated with (but not slaved to) the mass density (which is fully quantum).

It is important to emphasize that there is no fundamental justification for this fact, except for the plain desire to define a classical channel for the gravitational interaction.
The idea that Newtonian gravity can be described by a classical / stochastic interaction channel contradicts important {\em facts}  about the gravitational field. In particular, it contradicts our understanding that in weak gravity,  save for  gravitational waves  which are the true dynamical degrees of freedom, gravity is completely slaved to the distribution of matter. Any theory that violates this property must explain the physical origin of this violation. After all, Newtonian gravity is embedded in General Relativity, so any modification of the former implies invariably a modification in the later. In particular, the slaving of the potential to matter via Poisson equation is a constraint of General Relativity, which expresses the invariance of the theory under time reparameterizations, a fundamental symmetry of GR\footnote{Note that the KTM model and some of its variations are incompatible with atomic fountain data \cite{ACMZ}. However, our critique about incompatibility with GR applies to all possible implementations of classical channels for Newtonian gravity.}.

The motivation for proposals by KTM and TD is the idea that gravity, i.e., spacetime geometry, is fundamentally classical, and that it should not be quantized.  Along this way of thinking,
quantum matter coupled to classical gravity invariably involve stochastic behavior for the gravitational dynamical variables.  However,  if one agrees that classical gravity is governed by GR,  then at the weak gravity level, the only dynamical variables are gravitational waves; the Newtonian force is certainly not.  The assumption of a quantum- classical coupling of matter with  Newtonian gravity, which involves no true degrees of freedom for the gravitational field, is inexplicable from the perspective of GR.

In the language of information channels, GR predicts unambiguously a direct interaction channel between two separated quantum mass distributions, obtained by adding the quantum version of the operator $H_{gsi}$ in the Hamiltonian for quantum matter. This channel will lead to phenomena like entanglement generation \cite{Bose17, Vedral17} and  Rabi-type oscillations \cite{AnHu20}. However, since this channel is direct, the observation of such phenomena says nothing about the nature of the true degrees of freedom of the gravitational field. As the latter effectively decouple from the Newtonian interaction at the level of weak gravity, one cannot make any statement about their nature, classical or quantum, from the observation of gravity-induced entanglement generation.

The observation of gravity-induced entanglement certainly rules out models in which Newtonian gravity corresponds to a classical / stochastic channel, or models with a classical mediator for Newtonian gravity. However, as explained here, these models are rather implausible from the perspective of gravity theory. A more plausible class of models involves a classical / stochastic interaction mediated by the true degrees of freedom of gravity (gravitational waves), together with the direct-channel interaction of Newtonian gravity, in the vein of the Anastopoulos-Blencowe-Hu model of gravitational decoherence \cite{AHGravDec, Miles}. These models are fully compatible with gravity-induced entanglement in the Newtonian regime, and they provide an explicit counter-example to claims of proving the quantumness of gravity from experiments with Newtonian gravity.

 Ref. \cite{claims2} presents an argument why  a mass  in a spatial superposition that interacts gravitationally with a test mass leads either  to faster-than-light signalling or violation of quantum complementarity, unless one takes into account vacuum fluctuations of the gravitational field, and the emission of gravitational radiation. However, a close reading of the argument shows that vacuum fluctuations need not be quantum, and that the restoration of quantum complementarity only requires a decoherence mechanism---spontaneous emission of discrete quanta being only one of many  possible scenarios. Hence, the arguments of Ref. \cite{claims2} are not stringent enough to rule out theories in which the true degrees of freedom of gravity are treated as a classical stochastic field that causes decoherence to quantum matter.
But these are  properties that any mathematically consistent quantum-to-classical coupling must have anyway\footnote{
Mathematically consistent couplings of quantum to classical variables typically induce non-unitary evolution and decoherence to the quantum system and noise to the classical system \cite{BlJa, HaDi, Diosi3, HaRe05}.
Models for quantum-to-classical coupling without decoherence and noise typically lead to the violation of positivity, i.e., negative probabilities.}.

\subsection{Event formalism and closed timelike curves}

The event formalism \cite{event1, event2} is an alternative quantum theory that predicts photon disentanglement in the presence of gravity.  It is built under the assumption that the  operators representing observables at different points along a particle's geodesic are required to commute with one another. This requirement implies a strong and rather {\em ad-hoc} modification to the properties of quantum fields.
This formalism stems from the  analysis of \cite{Ralph1} in finding possible unitary solutions to a quantum gravity information paradox. It is based on the treatment in \cite{Deutsch} of closed timelike curves in the context of quantum computation.

We want to make two points in relation to these models. First, the modification stems from arguments that are based on quantum computation  in terms of quantum circuits,  not on a QFT analysis.  As \cite{Deutsch} asserts, such computational circuits are universal in the sense that they can simulate the behavior of finite quantum systems. This means that they can simulate the behavior of a {\em finite} number of modes of the quantum field. However, the simulation of a QFT can  be highly non-local, as, for example, when simulating fermion fields with qubits \cite{fequ}. There is no guarantee that the computational degrees of freedom are localized on spacetime, and that they are subject to the usual analysis of locality and causality.

More importantly, the idea that closed timelike curves exist without  accompanying quantum gravity effects is highly implausible from the perspective of gravity theory. This strongly contradicts the well-motivated  {\em chronology protection conjecture} \cite{HawkingCP} which asserts that the laws of physics, including quantum phenomena, do not allow for the appearance of closed time-like curves. Chronology protection is valid also for semi-classical gravity \cite{KRW}. All this  strongly suggests  that closed timelike curves can emerge only as Planck scale quantum gravity effects (like Wheeler's spacetime foam \cite{Wheeler}), if at all. Hence, any phenomena stemming from closed timelike curves stretch the capabilities of current experiments more than 20 orders of magnitude, so there is no surprise of null findings \cite{Xuetal}.


\section{Problem 2:  Quantum processes / fluctuations cannot be replaced by classical stochastic processes / noises}

In this and in the next section we turn our attention to the relation of quantum fluctuations and classical stochastic processes.  Noise is often introduced and stochastic equations invoked in many alternative quantum theories for some specific purposes, such as in explaining why we don't see quantum effects  in the macroscopic world (e.g., the stochastic Schr\"odinger equation in \cite{Diosi89,GGR90}).    We begin by showing   how fluctuations, often at the Planck scale,  have been proposed as  the source of gravitational decoherence, and what goes wrong in doing so.  Gravitational decoherence is a good case study as it involves all three aspects: gravity, quantum field and quantum information.

\subsection{Fluctuations as sources of gravitational decoherence -- what is missing or misleading}

In many proposed models, gravitational decoherence  is a consequence of fluctuations that are assumed to originate from Planck-scale physics. The specific mechanism varies. Fluctuations may be induced by  some fundamental imprecision in the measuring devices (starting with clocks and rulers) \cite{GPP}, or by uncertainties in the dynamics \cite{MilIntDec}, or by treating time as a statistical variable \cite{Bonifacio}, to give some examples.

Obviously, any model that involves Planck scale physics must make   very strong assumptions. However, we find assumptions that Planck scale uncertainties can be modeled by classical stochastic processes rather implausible.  The modeling of  uncertainties by classical noise may work for randomness at the macroscopic scale where  quantum properties do not play a role.  In contrast, quantum uncertainties are different in nature  as they involve non-localities and correlations with no analogues in the classical theory of stochastic processes.

To explain this point, we note that the limitations posed by the Planck length are not {\em a priori} different from those placed by the scale $\sqrt{\hbar/c^3} e/m_e$ in quantum electrodynamics, where $e$ is the electron charge and $m_e$ is the electron mass.  At this scale quantum field effects are strong, and the fluctuations from these effects are fully quantum. Any effect they
cause at low energy is also inherently quantum. One needs to  specify the conditions (e.g., Gaussian systems) or the regime for the quantum field,  and justify the means   by which they could be  treated like classical fluctuations described by a stochastic process. In particular, the effects of the fluctuations of the electromagnetic field at low energies ($E<<mc^2$) have been well studied. It has been shown that the `noise' induced by these fluctuations is non-Markovian and does not cause significant decoherence effects in the microscopic regime \cite{Emdec}.  In other words, the coherence of the electromagnetic field vacuum does not allow for the {\em a priori} generation of classical (i.e., decohering) fluctuations in the quantum motion of the particle.  The assumption that the gravitational field exhibits a different behavior is completely {\em ad hoc}, with no justification unless one postulates that {\em gravity is fundamentally classical}.

\subsection{Classical stochastic processes or noises miss out important  information in quantum theories}

When encountering the effects of quantum fluctuations, a  common practice of authors who know enough about classical stochastic processes is to assert or assume that they can be replaced by classical stochastic processes.  Without a rigorous proof,   this substitution is unwarranted. Making this jump could also be dangerous,  if one does not even know what has been left out or what can be messed up in this substitution.  Important quantum features are ignored which bear on basic quantum information issues such as entanglement and coherence.

In some cases,  quantum fluctuations of an observable may be replaced by a classical stochastic process {\em after coarse-graining}. This is possible if the coarse-grained observable satisfies specific consistency conditions  \cite{GeHa, Omn}. Then, it can be rigorously shown that the measurement outcomes for this observable can be  modeled by a stochastic process  \cite{Ana02}. To this end, it is necessary that all specific quantum properties of the fluctuations are suppressed by coarse-graining.

Hence, except for specific limiting regimes, quantum processes cannot be replaced by classical stochastic processes.
We shall show that even in cases of the closest proximity,  namely,  for Gaussian systems, where the Wigner function remains positive definite and has the same form as a classical distribution function,   there are subtle and important differences:  a quantum theory contains more information than the corresponding classical stochastic ones.



 The statistical properties of a classical stochastic field can be described by a classical distribution functional of the field variable and its conjugate momentum over an infinite-dimensional phase space spanned by the canonical pair.  The distribution function of the classical stochastic field is often misconstrued as the equivalent of the quantum Wigner function.  Even for Gaussian systems it is not completely true that its quantum behavior  is  identical to the classical:  Wigner functions carry more information than what is in the corresponding classical probability distributions. This is no surprise because Wigner function  carries the full equivalent information  contained in the density matrix.

Can a classical stochastic field   act as a  functional surrogate of a quantum field?  This old issue is revisited in a recent paper by Hsiang and Hu (HH)  \cite{HHCosEnt}, where a conduit for Gaussian systems is built  connecting a full quantum field-theoretic treatment and those using the probability distribution of classical stochastic fields.  Since the bridging protocol is the source of a great deal of confusion, HH examine  the conditions for the two theories to be connected, while paying special attention to the ability of the latter to  preserve essential quantum properties.
They conclude  that the information contained in {\it classical stochastic field theory is far from complete}, e.g., it still needs inputs from the two-point functions of the quantum field to yield the stochastic counterparts.
Even though from certain angles both look formally identical, the theoretical frameworks they are based on are fundamentally different.

We list the key points from their findings,  beginning with the obvious question: A)  How is {quantum non-commutativity} represented?    In quantum theory, the Wigner function corresponds to a {\it fully symmetrized ordering}.  HH demonstrate  that for those  physical quantities of interest which can be expressed in terms of the covariance matrix elements, which are the expectation values of the canonical operators in {\it symmetrized ordering},  the stochastic field approach seems to work.
 But this is unlikely for other ordering choices.  In the classical stochastic field approach, different distribution functions may be needed to account for different operator orderings in quantum theory.    B) The Wigner functions for the more general {\it non-Gaussian configurations}  are not  positive-definite and the correspondence with classical distribution functions breaks down.
Crucial for {\it quantum information issues}, C) classical field theory, including stochastic renditions,  cannot account for quantum entanglement,  in which the nature of quantum state plays an important role.

We can see a glimpse of this from D) the differences between the classical Shannon entropy and the quantum von Neumann entropy.   The Shannon entropy associated with  classical distributions has the property of monotonicity. That is, given the combined systems $A$ and $B$, we have $\mathcal{S}_{s}[A+B]\geq\max\{\mathcal{S}_{s}[A],\mathcal{S}_{s}[B]\}$ for the Shannon entropy, where $\mathcal{S}_{s}[A]$ denotes the Shannon entropy of the subsystem $A$. On the other hand, for the von Neumann entropy $\mathcal{S}$, the closest property to the classical monotonicity is the theorem of Araki--Lieb  triangle~inequality:
\begin{equation}
	\lvert \mathcal{S}[A]-\mathcal{S}[B]\rvert\leq \mathcal{S}[A+B]\leq \mathcal{S}[A]+\mathcal{S}[B]\,.
\end{equation}
The second group of inequalities is known as the subadditivity inequality for von Neumann entropy, and holds with equality if and only if systems $A$ and $B$ are uncorrelated, that is, $\varrho^{AB}=\varrho^{A}\otimes\varrho^{B}$.   HH has shown (in the case of particle creation, where subsystems $A, B$ represents the   $\pm {\bf k}$ modes of the pair)   that  $\mathcal{S}[A,B]=0$ but $\mathcal{S}[A]=\mathcal{S}[B]>0$. The~entropy of a subsystem is larger than the entropy of the combined system. These properties are broadly known, often invoked in the discussions of entanglement entropy.

The message we wish to convey is, as a matter of principle, one should seek a quantum field description where the effects of quantum fluctuations are fully incorporated.   And,  if  or when a    classical stochastic process may provide an adequate description of the quantum field, specify the conditions and spell out the limitations.

\section{Problem 3:  How are semiclassical and stochastic theories related to their quantum origins? How does noise enter?}

The second question raised here is easier to answer. If  one permits noise to be added by hand, there is always a stochastic counterpart to any theory.   This practice has turned into an unquestioned habit in many situations,  likely attributable to the way how we  first encounter a stochastic equation, such as the Langevin equation: ``just add a noise term as source driving the Newton equation."  Voila!  Only later did we learn that in doing so a closed system where the dynamics is unitary is changed to an open system following nonunitary dynamics. We need to identify an environment which our system interacts with and  find out the coarse-graining measures which produce the noise chosen,  which drives the system.  Furthermore, we need to make sure the system's dissipative dynamics and the environmental noise observe a self-consistency requirement, such as governed by a fluctuation-dissipation relation.  Many proposers of  alternative quantum theories like to introduce a noise source,  turning, say, a Schr\"odinger equation into a stochastic Schr\"odinger  equation, which serves their specific purposes.  Regrettably,  rarely do these authors  explain where their noises come from, their physical meanings, and whether turning a unitary equation into a  stochastic equation generates any mathematical conflicts with,  or compromises the theoretical foundation of,  the established  theories.

There is one situation we know of and have worked on, where a classical stochastic source (or noise) can be derived as a full representation of quantum field fluctuations, for  systems and environments being all Gaussian.  We shall explain this in the next subsection.

Let us now turn to the first question:  How are semiclassical and stochastic theories  related to their progenitor quantum theories?
For this discussion,  it is useful to keep in mind the four distinct levels of theoretical structures: quantum, stochastic,  semiclassical and  classical, in the  order of decreasing information contents.  Every lower level theory is a limiting case of the higher level theories, with the quantum theory as their source.  Note in particular the stochastic level here is different from theories obtained by putting a noise in by hand. Instead, stochastic follows quantum only because it is a genuine derivative of the quantum theory under specific conditions.   It is useful to review this four-level structure so that as one sees some stochastic source entering one can identify it in the right place and decide whether it can be derived or just  introduced in an ad hoc manner,  a big difference.

\subsection{Semiclassical theory as large $N$ limit of quantum theory}

The relation between semiclassical and quantum theory is a well-studied subject ever since quantum theory got established, and there is almost a semiclassical theory for every quantum theory which came into being, constructed from the corresponding longer existing and more familiar classical theory: quantum theory of radiation, quantum chaos, etc.  There are also cases where the quantum theory is well established but a lesser theory is introduced with the help of experimental input in order to bypass the difficult calculations, Lifshitz's stochastic electromagnetic theory is a good example, with monographs devoted to it \cite{StoEM}.  Semiclassical theory plays a special role for gravity because of the lack of a {\it bona fide} quantum theory of (nonperturbative) gravity.   There, gravity is kept classical, described  by Einstein's theory of general relativity (GR), while matter is described by  quantum field theory (QFT), both theories have been tested to the utmost degree, the best we have for the description of spacetime and matter.

Since there is a great deal of interest in semiclassical gravity theory among researchers of gravity, quantum fields and quantum information, and there are misconceptions of what a correct SCG theory is.  let us start with the popular Newton-Schrodinger equation.  On surface it looks like the weak field   limit of general relativity  and the nonrelativistic limit  of the Klein-Gordon or Dirac equation.  But it is not.  Single or few particle NSEq cannot be derived from GR+ QFT---see \cite{AHNSE}  for details or  \cite{HuAQT} for a summary.

Semiclassical theory makes sense as the large $N$ limit of the progenitor quantum theory.   The semiclassical Einstein equations, the Einstein tensor $G_{\mu \nu} = - 8\pi G \langle \hat T_{\mu \nu} \rangle,$ where $\hat T_{\mu \nu}$ is the stress energy tensor of quantum matter field,  are only meaningful for $N$-particle quantum states with $N > > 1$, where the relative strength of fluctuations is suppressed by a factor of $N^{-1/2}$.
Theories using the Newton-Schr\"odinger equation and theories that treat the  Einstein-Klein-Gordon,  Einstein-Dirac equations or the Moller-Rosenfeld equation \cite{MolRos}\footnote{Please refer to the four levels, from 0 to 3,  of ``semiclassical gravity' theories in \cite{HuAQT}.  The proper ones to use are Levels 2 and 3 (p. 7),  what are referred to as `relativistic semiclassical gravity' and `stochastic semiclassical gravity'.} in a {\it prima face}, abridged or altered way, are not compatible with GR+QFT. The crucial point is in the coupling between classical gravity and quantum fields,  such as stated in the Wald axioms \cite{Wald},  and the requirement of self-consistency.   The large $N$ limit of (perturbative) quantum gravity has been shown \cite{HarHor} to be   a pathology-free and self-consistent semiclassical gravity theory based on the relativistic semiclassical Einstein equations.

In the next subsection,  after we have introduced the stochastic semiclassical  gravity theory \cite{HVLivRev,HuVer20} based on the Einstein-Langevin equation \cite{ELE}, we shall see that the  relativistic semiclassical gravity theory  is obtained after performing  a stochastic average over the noise which describes the quantum fluctuations of matter field.  In this sense  relativistic semiclassical gravity is a  relativistic ``mean field" theory.


\subsection{Stochastic semiclassical theory:  noise can be defined for quantum fluctuations in Gaussian systems}

In the beginning of this section we mentioned that noise in a stochastic equation is often put in by hand. However,  one can define noise of a Gaussian quantum environment in a mathematically rigorous manner,  namely,  by way of the Feynman-Vernon \cite{FeyVer} Gaussian functional identity. Because of the way how this noise is derived, it is best to keep in mind its quantum origin, namely,  that it is only a stochastic classical source {\it representation} of quantum fluctuations, its true nature being fully quantum mechanical.  In this way one can distinguish from ad hoc noises attached  to a quantum equation such as in the stochastic Schr\"odinger equation,  or put in by hand to some  lower (classical or semiclassical) level equations.

Let us examine stochastic semiclassical gravity in the large $N$ vein.  When the quantum gravity sector is coupled to a large number $N$ of matter (free) fields, the lowest-order contribution in a $1/N$ expansion produces the semiclassical Einstein equations. When using large $N$ the cut-off scale is shifted to the rescaled length. At the lowest order the fluctuations of the
metric are suppressed but all matter loops are included.
In the same vein, stochastic gravity results from taking the next-to-leading
order in the $1/N$ expansion of quantum gravity coupled to $N$ matter fields.
Whereas in semiclassical gravity the fluctuations of the metric are suppressed,
here in stochastic gravity,  metric fluctuations are of linear order,  while graviton loops or internal graviton lines are sub-leading in comparison to matter loops. Semiclassical and stochastic gravity as effective theories in the large $N$ context are discussed in
Chapters 9 and 10  of \cite{HuVer20}.

Harking back to the issue of gravitational decoherence  by stochastic processes at the Planck scale, we remarked that models which introduce noise or invoke fluctuations which are  intrinsically classical have pathologies.  The proper way to introduce stochasticity in spacetime which respects both general relativity and quantum field theory is through the metric fluctuations induced by the fluctuations of quantum matter fields (via the noise kernel which is the correlator of the stress energy tensor of quantum matter field, see, e.g., \cite{PH01,OsbSho}) by seeking solutions to the Einstein-Langevin equation (ELE) \cite{ELE}. This has been carried out successfully for the Minkowski metric in \cite{MarVer}, and for inflationary cosmology in \cite{RouVer}.  In stochastic semiclassical  gravity the noise is fundamentally quantum;  the backreaction of a quantum matter field and its fluctuations on the classical geometry and its fluctuations is obtained by solving  the ELE which guarantees fully the self-consistency between the quantum matter field and the classical geometry sectors. In fact, the magnitude of metric fluctuations enters as the most stringent criterion of the validity of semiclassical gravity  \cite{FlaWal,HRValidSCG}.

Returning now  to the low energy limit, where laboratory or space experiments can provide some reality checks, upgrade to a stochastic Newton-Schr\"odinger equation has been proposed \cite{Bera} with the intention of explaining the origin of noise in some of the alternative quantum theories.  Here again, our critiques of the NSE apply.  The proper way is to take the weak field nonrelativistic limit of the Einstein-Langevin equation.   This limit of stochastic gravity is of  special interest as it is the only known consistent theory which can deal with quantum entanglement issues in a gravitational setting, such as in the physics of gravitational cat states \cite{AnHu20}  (See also Sec. 1.3.4 of \cite{HuVer20}.)

In summary,  for the points raised in this section,  in the consideration of  gravity and quantum fields,  back-reaction self-consistency is the ultimate criterion to check on:  from the  way how these two sectors are coupled, the way how noise of quantum field is derived, how the quantum  expectation values and  stochastic averages are taken, how  a quantum field and its fluctuations back-react on a classical background spacetime in a self-consistent manner,  are all of crucial importance. Improper treatments or assumptions in the consideration of any of these factors and their related issues can lead to pathologies disallowed in general relativity and quantum field theory.\\

\noindent {\bf Acknowledgment} This work is supported in part by a Schwinger foundation grant JSF-19-07- 0001.

\end{document}